\documentclass[a4paper]{jpconf}
\usepackage{graphicx}
\usepackage{subfig}
\begin{document}
\title{Beam Normal Single Spin Asymmetry Measurements from Q$_{weak}$}

\author{Buddhini P. Waidyawansa for the Q$_{weak}$ Collaboration}

\address{C122, 12000 Jefferson Avenue, Newport News, VA 23602}

\ead{buddhini@jlab.org}

\begin{abstract} 
The Q$_{weak}$ experiment has made several interesting beam normal single spin asymmetry measurements. Preliminary result from a 3.2\% measurement of the beam normal single spin asymmetry in elastic e+p scattering at E = 1.155 GeV and $\theta_{lab} = 7.8^\circ$ is presented. We have also made measurements of this asymmetry in elastic and inelastic scattering in the Delta resonance region from Hydrogen, Aluminum and Carbon targets and e+e scattering from Hydrogen target. Some initial results from these measurements are also presented. 
\end{abstract}

\section{Introduction}

A beam normal single spin asymmetry ($B_n$) is measured in the scattering of transversely polarized electrons from unpolarized nucleons. The asymmetry arises from the interference between the one-photon exchange and the two-photon exchange processes. $B_n$ is a parity-violating and time-reversal invariant observable of the imaginary part of the two-photon exchange process and can be written as \cite{DeRujula:1972te}
\begin{equation}
\label{eq:B_n_theory}
B_n = \frac{\mathcal{I}m(T_{1\gamma}\bullet AbsT_{2\gamma})}{|T_{1\gamma}|^2}.
\end{equation}
Here $T_{1\gamma}$ is the scattering amplitude for the one-photon exchange process, $\mathcal{I}m$ denotes the imaginary part, and $AbsT_{2\gamma}$ is the absorptive part of the two-photon exchange amplitude. Over the years, measurements of $B_n$ have become feasible using apparatus designed to make few \% measurements of asymmetries such as the Q$_{weak}$ apparatus.



\section{Experimental Setup}

The experiment took place in experimental Hall C at the Thomas Jefferson National Accelerator Facility in Virgina. A 150-180 $\mu A$ polarized electron beam was generated at the injector and was accelerated to an energy of 1.16 GeV before reaching the target inside the Q$_{weak}$ setup located in Hall C. The electron polarization direction was adjusted using two Wien filters located in the injector. Data were collected using two transverse polarization configurations: vertical (spin pointing up) and horizontal (spin pointing to beam left at the target). In each configuration, the helicity of the electrons were flipped between +(spin up/spin left) and - (spin down/spin right) at 960 Hz in a pseudo-random manner with the use of a pockels cell located in the injector. The electrons were scattered from the target and electrons with an angle of 5.8$^{\circ}$-11.0$^{\circ}$ were selected using a set of collimators. The energy of the scattered electrons were 1.155 $\pm$ 0.003 GeV. A toroidal magnet was then used to focus the elastic electrons onto an array of eight Cerenkov detectors located symmetrically around the beamline with $\approx$ 50\% azimuthal acceptance. A detailed description of the setup can be found in \cite{Allison:2014tpu}.


\section{Analysis Overview}

Detector asymmetries were formed using $A_{det}=(Y^+ - Y^-)/(Y^+ + Y^-)$ where Y$^{\pm}$ is the detector yield in the $\pm$ helicity state. The helicity correlated false asymmetries in $A_{det}$ generated by beam parameters were largely removed during the experiment by the use of an insertable half wave plate at the injector which was used to flip the electron helicity periodically ($\approx$ 8 hrs) independently of the helicity signal used to control the pockels cell. The remaining helicity correlated beam asymmetries were removed during the analysis stage using multiple linear regression. 

\begin{figure}[!htbp]
\centering
\subfloat[][]{\includegraphics[width=0.5\textwidth]{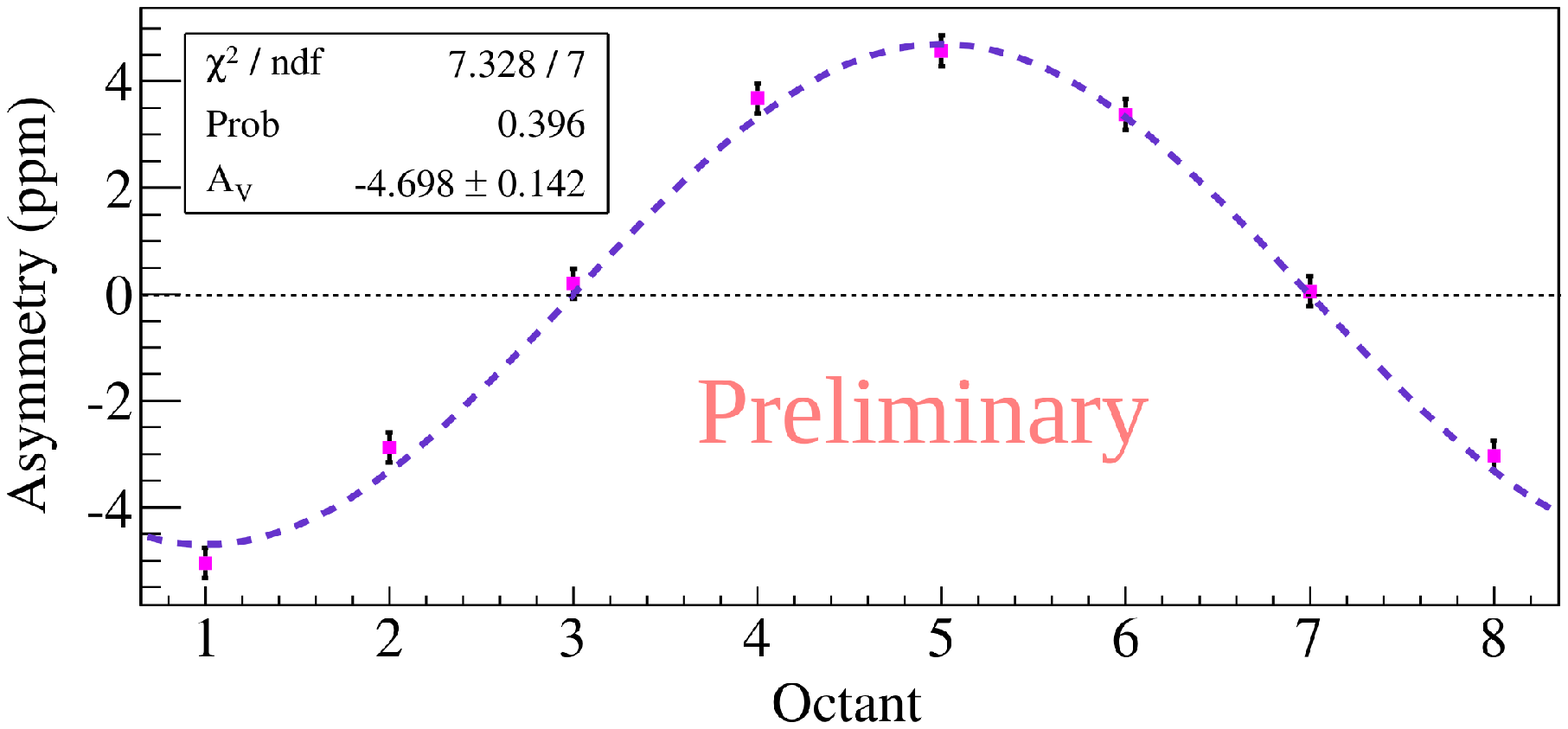}}
\subfloat[][]{\includegraphics[width=0.5\textwidth]{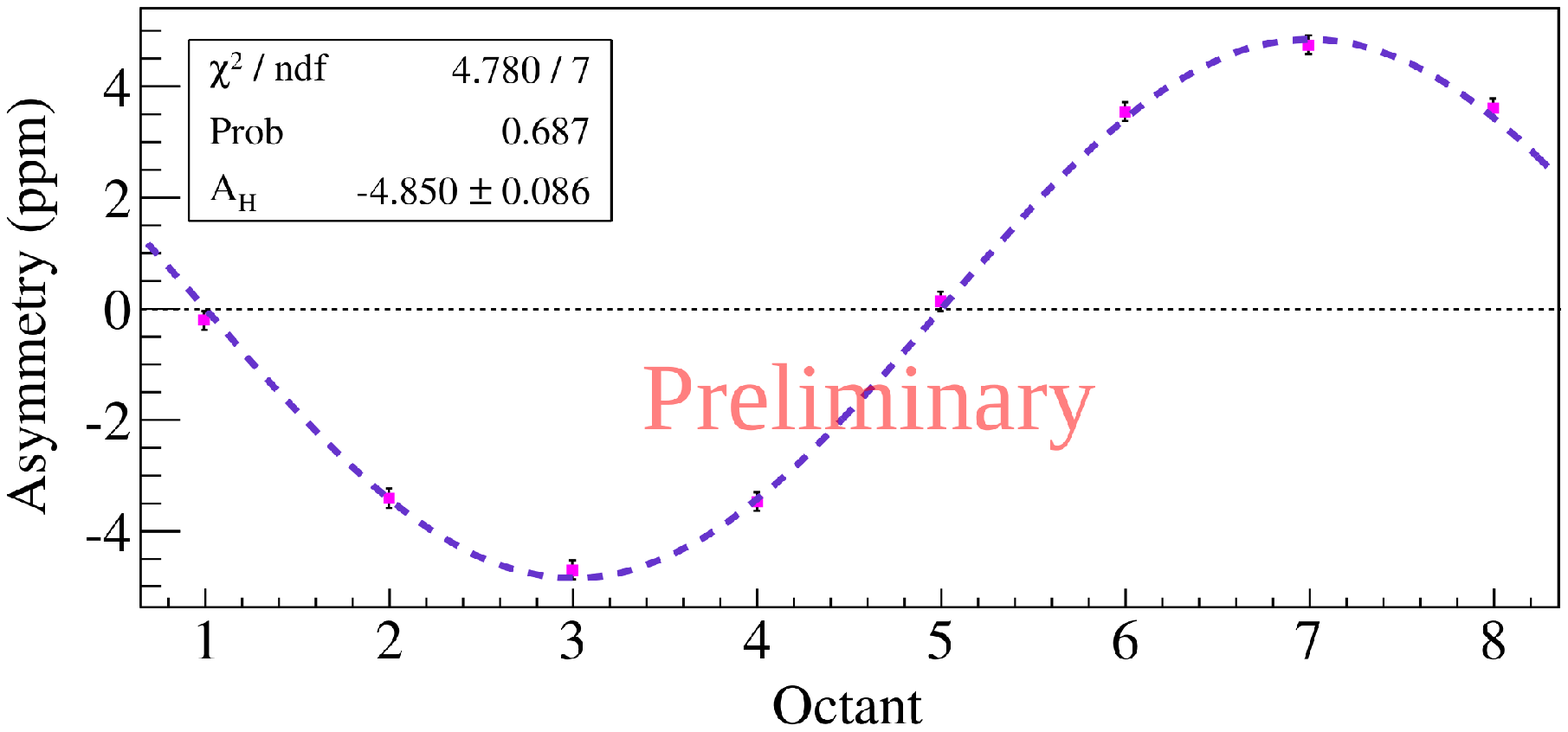}}
\caption[]{\label{fig:fit-representation}
Representation of the fitting procedure used to extract the experimental asymmetry from (a) vertical transverse ($\phi_s = 90^{\circ}$) and (b) horizontal transverse ($\phi_s = 0^{\circ}$) configurations. The data shown here are elastic e+p scattering from LH$_2$ target. The octant number corresponds to the main detector angle in the azimuthal plane with $\phi = 0^{\circ}$ at beam left and going clockwise. }%
\label{fig:cont}%
\end{figure}

For $B_n$, the asymmetry measured in a detector placed at an azimuthal angle $\phi_{det}$ has the form 
\begin{equation}
\label{eq:fit}
A_{det}(\phi_{det}) = A_{exp}\sin(\phi_{det} - \phi_s),
\end{equation}
where $\phi_{s}$ is the angle of the electron spin vector and $A_{exp}$ is the experimental asymmetry containing $B_n$, beam polarization, and backgrounds. By fitting the octant dependance of the detector asymmetries (corrected for insertable half wave flip) using Equation \ref{eq:fit} allows the extraction of $A_{exp}$ (see Figure \ref{fig:fit-representation}) . The experimental asymmetry extracted from the fit is then corrected for beam polarization and backgrounds with the use of 
\begin{equation}
\label{eq:bn}
B_n = R\left(\frac{A_{exp}/P - \sum^2_i A_{bkg}^if_i}{1-\sum_i^2f_i}\right)
\end{equation}
where R is a factor accounting for radiative corrections and kinematics, $f_i$ is the dilution of the i$^{th}$ background $A_{bkg}^i$ and P is the beam polarization. For the details of the full analysis see \cite{BWaidyawansa_phd}.

\newpage
\section{Preliminary Results from Elastic e+p Scattering}

50 hrs worth of elastic e+p events were collected from a liquid hydrogen target (LH$_2$).  Using Equation \ref{eq:bn}, the experimental asymmetry was corrected for beam polarization, kinematics, and backgrounds coming from the aluminum target windows and a few percent inelastic electrons. Both of these background corrections used dedicated asymmetry measurements which are discussed in the following sections. The resulting $B_n$ at E=1.155 GeV and $\theta=7.9^{\circ}$ is -5.35 $\pm$ 0.07 (stat) $\pm$ 0.15 (systematics) ppm.  This is a preliminary 3.2\% measurement. 

\begin{figure}[!htbp]
\centering
\includegraphics[scale=0.5]{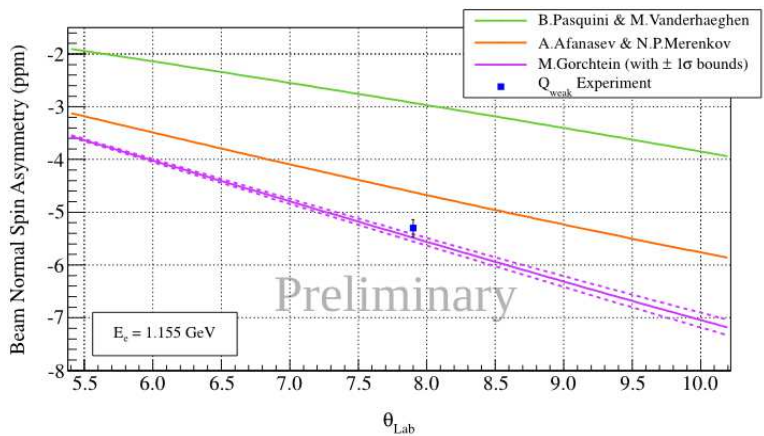}
\caption{\label{fig:elastic_ep}Comparison between our measurement of $B_n$ from elastic e+p scattering and theoretical calculation from B. Pasquini \cite{Pasquini:2004pv}, A.Afanasev \cite{Afanasev:2004pu} and M. Gorchtein \cite{Gorchtein:2005yz} within the acceptance of the experiment.}
\end{figure}

Figure \ref{fig:elastic_ep} shows a comparison between our measurement and several theory predictions. The Pasquini model is a calculation in the resonance region which use electroproduction amplitudes (MAID) to calculate the two-photon exchange contribution in $B_n$. In the resonance region they are limited to the use of intermediate states of the nucleon with single pion excitations. The Afansev and Gorchtein models use the Optical theorem and photo-production cross-sections to calculate the two-photon exchange amplitude. Their calculation uses multi-pion excitations in the intermediate nucleon state. Therefore, our measurement indicates that at forward angles, $B_n$ favors models with multi-pion excitations in the intermediate nucleon above the two-pion threshold. The differences between Afanasev and Gorchtein models, which are similar to one another except in parametrizations, is clearly visible due to the precision of our measurement.

\section{Initial Results from Elastic e+Al and e+C Scattering}
\begin{figure}[!htbp]
\centering
\includegraphics[scale=0.6]{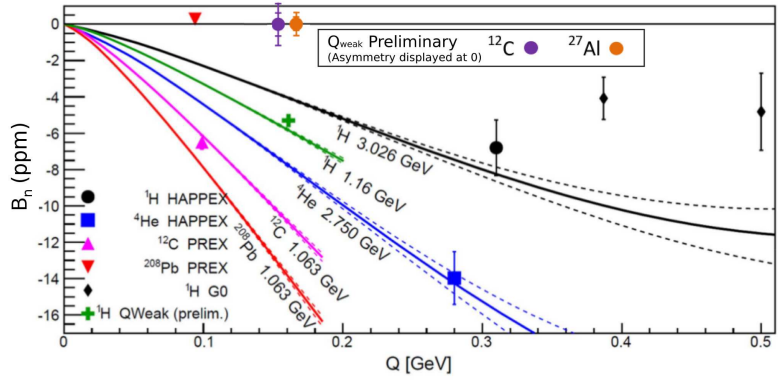}
\caption{\label{fig:al_theory}Comparison between experiments and theory predictions \cite{Gorchtein:2008dy} for $B_n$ from different targets. Published PREX and HAPPEX data are from \cite{Abrahamyan:2012cg}. Original figure from \cite{Dalton:2013cna}. Our measurements are shown at $B_n$ = 0 with statistical errors (inner error bars) and conservative systematic errors (outer error bars) for a qualitative comparison. }
\end{figure}

Model calculations \cite{Gorchtein:2008dy} of $B_n$ from nuclear targets have shown good agreement with existing data except in the case of lead (Pb) (see Figure \ref{fig:al_theory}). This difference between the theory and data may indicate a drastic change in the model between mass number A = 12 and A = 208 which may be verified with $B_n$ measurements on additional targets with A$>$12. 
 
\begin{figure}[!htbp]
\centering
\includegraphics[scale=0.35]{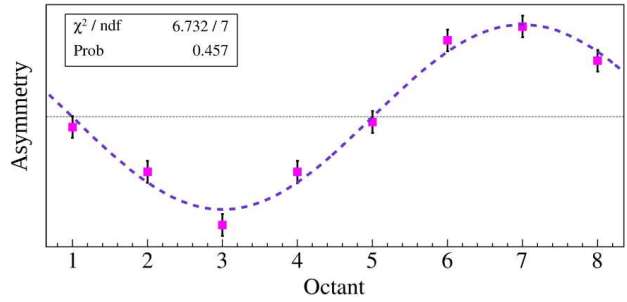}
\caption{\label{fig:al_fit}Octant fit of the elastic e+Al asymmetries measured with horizontal transverse beam polarization. The vertical axis is omitted until final publication.}
\end{figure}

We have measurements of $B_n$ from Al and C at E= 1.16 GeV and $Q^2=0.025$ GeV/C$^2$. The analysis of these data are still ongoing and are not at a stage where they can be directly compared to theory calculations. However, we can give an indication of the quality of these measurements with the use of statistical errors and conservative systematic errors as shown in Figure \ref{fig:al_theory}. In addition, Figure \ref{fig:al_fit} shows the octant fit over the Al asymmetries with a non-zero amplitude/experimental asymmetry. This observation is important as it now provides a data point to constrain the modeled behavior of $B_n$ from spin-0 nuclei between the mass number A = 12 and A = 208. 

For the completion of this analysis we require theoretical inputs on inelastic scattering from other nuclear excited states of Al and C (other than the $\Delta$) and the contributions from the contaminations ( ∼ 90\% Al, 5.7\% Zn, 2.5\% Mg,1.7\% Cu, ..) in the aluminum alloy target used for the measurements. Therefore, the progress of the analysis will benefit heavily from the inputs of the theoretical community.

\section{Initial Results from Inelastic Scattering in the $\Delta$ Resonance Region}

\begin{figure}[t]
\centering
\includegraphics[scale=0.35]{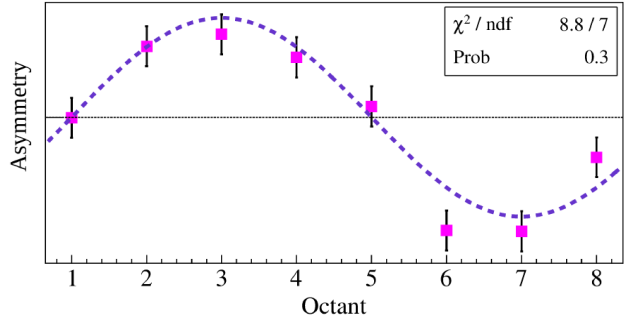}
\caption{\label{fig:inelastic_fit}Octant fit of the inelastic e+p asymmetries measured with horizontal transverse beam polarization. The vertical axis is omitted until final publication.}
\end{figure}

Measurements of $B_n$ from inelastic electron scattering is a unique tool to study the $\gamma \Delta\Delta$ form factors in two-photon exchange process with a $\Delta$ in the final state. Although incomplete \cite{MarkV14}, the theory predictions for $B_n$ at the $\Delta$ resonance region shows the expected asymmetry at very forward angles is positive and large \cite{Pasquini09}. Our analysis of the inelastic asymmetries are currently not at a stage where we can directly compare them to theory. However, an initial look at the octant fit (see Figure \ref{fig:inelastic_fit}) of the LH$_2$ data shows the experimental asymmetry is non-zero and is opposite in phase to the elastic asymmetries shown in Figure \ref{fig:elastic_ep}. 

\newpage
\section{Elastic e+e Scattering}
\begin{figure}[!htbp]
\centering
\includegraphics[scale=0.4]{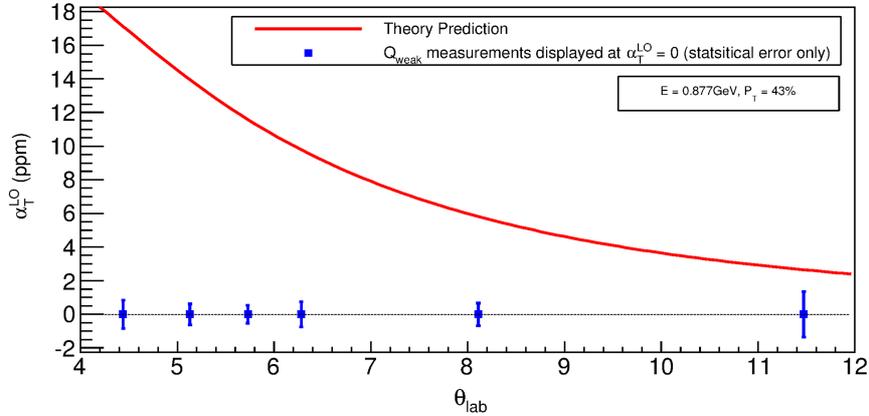}
\caption{\label{fig:moller_trans}Theory predictions \cite{Dixon:2004qg} for $B_n$ from e+e scattering around the Moller peak. The vertical axis gives the leading order asymmetry ($\alpha_{T}^{LO}$) and the horizontal axis gives the scattering angle in the laboratory frame. The experimental points are displayed at A=0 with statistical error bars for a qualitative representation of the data.}
\end{figure}

As a result of a systematic measurement carried out with a 43\% transversely polarized beam, we have measured $B_n$ from elastic e+e scattering at E = 0.877 GeV. The data were collected at various kinematic points centered around the Moller peak by varying the field of the toroidal spectrometer. According to the model calculation of \cite{Dixon:2004qg}, in our kinematics, the asymmetry is positive and in the order of few ppm (see Figure \ref{fig:moller_trans}). The analysis of the data is ongoing. But we have included the statistical precisions of each data point in Figure \ref{fig:moller_trans} to give an indication of what can be expected from the final results, assuming the systematics are well under control.

\section{Summary}
Q$_{weak}$ has new measurements of $B_n$ from several targets (H, Al, C) at several interesting kinematic regimes (elastic, inelastic and Moller). The analysis of the data set corresponding to elastic scattering from hydrogen is complete. This resulted in a preliminary 3.2\% measurement which is expected to improve further with refining of systematic errors. The analysis of the rest of the data are ongoing and are expected to be completed within the next few years with the help of theoretical inputs.   

\section{References}
\bibliography{pavi14.bib}

\end{document}